\documentclass[12pt,prd,aps,amssymb,amsmath,tightenlines,showpacs,a4paper,nofootinbib]{revtex4}
\def\bea{\begin{eqnarray}}
\def\eea{\end{eqnarray}}
\def\beax{\begin{eqnarray*}}
\def\eeax{\end{eqnarray*}}
\def\half{\frac{1}{2}}

\def\vf{\varphi}

\begin{document}
\title{Singular Mapping for a $PT$-Symmetric Sinusoidal Optical Lattice at the Symmetry-Breaking Threshold}

\author{H.~F.~Jones$^\dag$\email{h.f.jones@imperial.ac.uk}}

\affiliation{
$\phantom{.}^\dag$Physics Department, Imperial College, London SW7 2BZ, UK\\}
%\date{today}

\begin{abstract}
 A popular $PT$-symmetric optical potential (variation of the refractive index) that supports a variety of interesting and unusual phenomena is the imaginary exponential, the limiting case of the potential $V_0[\cos(2\pi x/a)+i\lambda\sin(2\pi x/a)]$ as $\lambda \to 1$, the symmetry-breaking point. For $\lambda<1$, when the spectrum is entirely real, there is a well-known mapping by a similarity transformation to an equivalent Hermitian potential. However, as $\lambda \to 1$, the spectrum, while remaining real, contains Jordan blocks in which eigenvalues and the corresponding eigenfunctions coincide. In this limit the similarity transformation becomes singular. Nonetheless, we show that the mapping from the original potential to its Hermitian counterpart can still be implemented; however, the inverse mapping breaks down. We also illuminate the role of Jordan associated functions in the original problem, showing that they map onto eigenfunctions in the associated Hermitian problem.

\end{abstract}

\pacs{03.65.Ca, 42.25.Bs, 42.25.Bs, 11.30.Er, 02.30.Gp}
\maketitle
%%%%%%%%%%%%%%%%%%%%%%%%%%%%%%%%%%%%%%%%%%%%%%%%%%%%
\section{Introduction}

The idea of $PT$ symmetry in the context of quantum mechanics began with the paper of Bender and Boettcher\cite{BB}, which showed both numerically and using the WKB approximation that the eigenvalues of the class of Hamiltonians $H=p^2-(ix)^N$, for $N\ge 2$, were all real and positive, in spite of the fact that these Hamiltonians are not Hermitian. Instead, they are invariant under the combination of $P$, whereby $x \to -x$ and $T$, whereby $i\to -i$. A rigorous proof of the reality of the eigenvalues came several years later in a paper by Dorey et al.\cite{DDT}, using ideas such as the Bethe ansatz and the ODE/IM correspondence between ordinary differential equation and integrable models.

It was then realized \cite{AM} that such Hamiltonians, possessing a completely real spectrum, could be mapped by a similarity transformation $\rho$ to an equivalent isospectral Hamilton $h$, according to $h=\rho H \rho^{-1}$. However, this transformation may be problematic \cite{BK, DK}, in that the operator $\rho$ or its inverse $\rho^{-1}$ may be unbounded. This is certainly the case when the spectrum of $H$ contains one or more Jordan blocks, where two eigenvalues, together with their eigenfunctions, coalesce. Typically this occurs at a critical value for a parameter in $H$, below which the spectrum is completely real, and above which complex-conjugate pairs of eigenvalues emerge, so that the $PT$ symmetry of the Hamiltonian is not respected by the eigenfunctions.

A very simple case in point is the potential $V=V_0[\cos(2\pi x/a)+i\lambda\sin(2\pi x/a)]$, where the $PT$ symmetry is respected for $\lambda<1$, but broken for $\lambda>1$. At the critical value $\lambda=1$, when $V=V_0 e^{2i\pi x/a}$, Jordan blocks appear, and the operator $\rho$ becomes singular. This particular potential has been the subject of intensive study in recent years, particularly in the context of classical optics \cite{op1}-\cite{HFJ1}, since the pioneering paper \cite{op1}, which pointed out the correspondence between the Schr\"odinger equation and the equation for the propagation of light in the paraxial approximation. In the optics context the role of $V$ is taken over by variations in the refractive index, with the imaginary part corresponding to gain and loss, which, in a $PT$-symmetric system, are delicately balanced.

In the case $\lambda<1$ the similarity transformation has been used \cite{MRC, HFJ2} to calculate the optical characteristics of $H$ using known properties of $h$. However, as noted above, this cannot be done in the critical case $\lambda=1$ because the transformation becomes singular. Nonetheless, in the present short note we show that some aspects of the similarity transformation still remain in the limit as $\lambda\to 1$. Specifically we show how the limit can be taken of the transformation from $H$ and its eigenfunctions to those of $h$. In particular this calculation elucidates the role of the Jordan associated functions that occur for $H$, showing how they map into eigenfunctions of $h$. However, the inverse transformation, from the simple eigenfunctions of $h$ to those of $H$, cannot be implemented.

%%%%%%%%%%%%%%%%%%%%%%%%%%%%%%%%%%%%%%%%%%%%%%%%%%%%
\section{The Similarity Transformation}
For simplicity we scale $x$ so that $a=\pi$, and choose $V_0=2$, so that the Hamiltonian reads
\bea\label{HH}
H = p^2-2(\cos{2x} + i \lambda\sin{2x})
\eea
Below the threshold for $PT$-symmetry breaking, $\lambda<1$, the real and imaginary parts of the potential can be combined into a cosine of complex argument, according to:
\beax
\cos{2x} + i\lambda\sin{2x}=\surd{(1-\lambda^2)}\cos(2x-i\theta),
\eeax
where $\theta={\rm arctanh}(\lambda)$. Thus $H$
can be converted into the equivalent Hermitian Hamiltonian
\bea\label{h}
h=p^2-2\surd{(1-\lambda^2)}\cos{2x}
\eea
by the complex shift $x\to x+\half i \theta$. This can be implemented by the similarity transformation
\bea
h=e^{-\half Q}H\ e^{\half Q}
\eea
with $Q=\theta \hat{p}\equiv -i \theta d/dx$, which ensures that the spectra of the two Hamiltonians are identical.

In the limit $\lambda \to 1$ the equivalent Hermitian Hamiltonian, $h$, becomes just the free Hamiltonian $h=p^2$, with eigenvalues $k^2$ and eigenfunctions $\vf_k(x)=e^{ikx}$.

For $\lambda=1$ we know the corresponding Bloch eigenfunctions of $H$, namely \cite{HFJ1}
\bea
\psi_k(x)=I_k(\sqrt{2}e^{ix}),
\eea
where $I_k$ is an associated Bessel function. However, when $k$ is an integer $n$, the eigenfunctions for $n$ and $-n$ are degenerate: $I_n(z)=I_{-n}(z)$, so that the spectrum has a Jordan-block structure. The eigenfunctions are no longer complete, and need to be supplemented by the associated Jordan functions (generalized eigenvalues) $\chi_n(x)$, satisfying
\bea
(H-n^2)\chi_n(x)=\psi_n(x)
\eea

Because $H$ is non-Hermitian, the eigenfunctions are not orthonormal in the usual sense. Instead the relevant overlap integral is the $PT$-overlap
\bea\label{PTmetric}
\int \psi_{-k}(x)\psi_{k'}(x)dx =\propto\delta_{kk'}
\eea
The degenerate eigenfunctions are self-orthonormal, but have a non-vanishing overlap with the corresponding Jordan associated functions.

As $\lambda$ approaches 1 from below, $\theta \to \infty$, so that the transformation becomes singular. Let us now explore the mapping from the eigenfunctions $\psi_k(x)$ of $H$ to the corresponding eigenfunctions $\vf_k(x)$ of $h$ in that limit. The relevant formula is the behaviour of $I_k(z)$ as $z\to 0$, namely (Eq.~(9.6.7) of Ref.~\cite{AS})
\bea\label{eq9.6.7}
I_k(z) \sim (z/2)^k/\Gamma(k+1)
\eea
for $k \ne -n$, a negative integer. The formula is relevant because when we make the substitution $x \to x+\half i \theta$  the argument $z=
\sqrt{2}e^{ix}$ of the Bessel function solutions for $H$ becomes $\sqrt{2}e^{-\half \theta} e^{ix}$, with $\theta\to \infty$.
Thus, for $k\ne -n$,
\bea\label{lim}
\psi_k(x)\equiv I_k(\sqrt{2} e^{ix}) \to \left(\half e^{-\theta}\right)^{\half k}\ \frac{e^{ikx}}{\Gamma(k+1)}
\eea
reproducing the expected eigenfunctions $\vf_k(x) \equiv e^{ikx}$, albeit with a prefactor that tends to zero for $k>0$, and to infinity for $k<0$.

On the other hand, the inverse transformation certainly does not work, because to reproduce $I_k(\sqrt{2} e^{ix})$ one needs all the terms in the series (Eq.~(9.6.10) of \cite{AS})
\bea
I_k(z)=\frac{(z/2)^k}{\Gamma(k+1)}\sum_{r=0}^\infty \frac{(\half z)^{2r}}{r!\Gamma(k+r+1)},
\eea
whereas in going from $I_k(\sqrt{2} e^{ix})$ to the $e^{ikx}$ using Eq.~(\ref{eq9.6.7}) the terms with $r>1$ are subdominant.

Let us now turn to case when $k=-n$, a negative integer. In that case we may not use Eq.~(\ref{eq9.6.7}) to obtain $\vf_{-n}(x)$, although we may still obtain $\vf_n(x)$ from $I_n(\sqrt{2} e^{ix})$. Correspondingly there is only one eigenvector in the non-Hermitian problem since $I_{-n}(z) =I_n(z)$. However, we must include the Jordan associated functions $\chi_n(x)$ in order to have a complete set of states.

Let us first look at the case $n=1$, to see how things work. In that case it was found in Ref.~\cite{GJ} that
\bea\label{chi1}
\chi_1(x)=-\frac{I_0(z)}{2z},
\eea
where again $z=\sqrt{2}e^{ix}$. Making the transformation from $x$ to $x-\half i \theta$ we obtain
\bea
\chi_1(x)\to -\frac{I_0(\sqrt{2}e^{ix-\half\theta})}{2\sqrt{2}e^{ix-\half\theta}}\sim \frac{e^{\half\theta}}{2\sqrt{2}}e^{-ix}
\eea
Thus it is the Jordan associated function that goes over to the missing eigenfunction $e^{-ix}$ of $h$. The factor $e^{-ix}$ comes from the denominator in Eq.~(\ref{chi1}).

The general case can be established by the definition of $\chi_n(x)$ as proportional to the derivative of the eigenfunction $I_k(z)$ with respect to $k$ evaluated at $k=n$, modulo solutions of the homogeneous equation $(H-n^2)\chi_n(x)=0$. The general formula (Eq.~9.6.44) of \cite{AS}) is
\bea
(-1)^n\frac{\partial}{\partial k}I_k(z)\Big|_{k=n}=-K_n(z)+\half n!\ \left(\frac{2}{z}\right)^n\ \sum_{r=0}^{n-1} (-1)^r\frac{(\half z)^r I_r(z)}{r!(n-r)!},
\eea
from which we must exclude the term in $K_n(z)$ in defining $\chi_n(z)$ in order to get the correct periodic behaviour.
When we now let $z\to \sqrt{2}e^{ix-\half\theta}$, subsequent terms in the series are subdominant compared with the first. Thus the $x$-behaviour of the transformed Jordan associated function arises solely from the prefactor, and is $e^{-ixn}$, reproducing the missing eigenfunction $\vf_{-n}(x)$

The reason that an associated Jordan function of $H$ goes over to an eigenfunction of $h$ is that when the equation $(H-n^2)\chi_n(x)=\psi_n(x)$ is transformed, the right-hand side acquires a factor of $e^{-\half n\theta}$, while the left-hand side
acquires a factor of $e^{\half n\theta}$. Thus in the limit the transformed equation becomes $(h-n^2)\vf_{-n}(x)=0$.

We may ask how the overlap integral for the $\psi_k(x)$, as given in Eq.~\ref{PTmetric},  maps into the standard overlap integral for the $\vf_k(x)$. In fact it maps over smoothly. Thus, in making the transformation (\ref{lim}) the factors $e^{\half k \theta}$ and  $e^{-\half k \theta}$ cancel out, and there is no problem with the limit as $\theta \to 0$ for $k\ne n$. In the case $k=n$, it is instead $\psi_n(x)$ and $\chi_n(x)$ that are orthogonal, and again the corresponding factors cancel out.

\section{Conclusion}

We have shown that in this singular limiting case the similarity transformation can still be used to map from the non-Hermitian problem to the equivalent Hermitian problem. It is interesting to note that where the Hermitian problem has a Jordan block structure with degenerate eigenfunctions, the corresponding eigenfunctions in the Hermitian problem are obtained by transforming the eigenfunction and its associated Jordan function. In the transformation appear very large or very small prefactors; however, these cancel out in the overlap integral. Unfortunately the mapping cannot be used in the reverse direction to derive the eigenfunctions and Jordan associated functions of $H$ from  the eigenfunctions of the free Hermitian Hamiltonian $h$.

%%%%%%%%%%%%%%%%%%%%%%%%%%%%%%%%%%%%%%%%%

\end{document}